\documentclass{article}
\usepackage{graphicx}
\usepackage{balance}  
\usepackage{authblk}
\newtheorem{theorem}{Theorem}
\usepackage{amsmath}

\begin{document}

\title{Efficient and Private Approximations of Distributed Databases Calculations}

\author[*]{Philip Derbeko Shlomi Dolev Ehud Gudes} 
\author[**]{\\ Jeffrey D. Ullman}

\affil[*]{philip.derbeko@gmail.com, dolev@cs.bgu.ac.il, ehud@cs.bgu.ac.il Ben-Gurion University of the Negev, Computer Science Department, Beer-Sheva, Israel}
\affil[**]{ullman@gmail.com Stanford University, Stanford, CA, USA}      

\maketitle

\begin{abstract}
\label{abstract}
In recent years, an increasing amount of data is collected in different and often, not cooperative, data\-bases. The problem of privacy-preserving, distributed calculations over separated data\-bases and, a relative to it, issue of private data release were intensively investigated.
However, despite a considerable progress, computational complexity, due to an increasing size of data, remains a limiting factor in real-world deployments, especially in case of privacy-preserving computations.

In this paper, we present a general method for trade off between performance and accuracy of distributed calculations by performing data sampling. Sampling was a topic of extensive research that recently received a boost of interest. 
We provide a sampling method targeted at separate, non-collaborating, vertically partitioned data\-sets. The method is exemplified and tested on approximation of intersection set both without and with privacy-preserving mechanism.
An analysis of the bound on error as a function of the sample size is discussed and heuristic algorithm is suggested to further improve the performance. The algorithms were implemented and experimental results confirm the validity of the approach. 
\end{abstract}

\section{Introduction}
\label{sec:introduction}
Consider different data providers holding vertically partitioned data. 
Each data provider holds different information about the same set of individuals and there is a common identifier (social security number, or any other sort of ID) that allows one to cross-reference individuals across data providers.
For example, the data providers contain a data of a set of individuals: gene ``banks'' hold genetic information, police departments hold criminal records, financial institutions keep record of credit history,  hospitals record health history of patience for future diagnosis, etc. 
Those data providers might be geographically spread, belong to different organizations and have their specific privacy and security requirements. For instance, a data provider of genetic information might not allow a public access to its data while allowing access for medical doctors to their patients data, or police departments that allow data access for international law enforcements, but deny access to anyone outside the departments. 
In such cases, it is usually impossible to gather all the data in one place, either due to the size of the data or due to privacy restrictions on data sharing. 
A client issuing a query to a different data providers in public cloud settings, might be required to pay for utilized CPU time and network traffic. In many cases, such client will be willing to trade performance (in other words, cost) for accuracy of the answer. 
Especially, if the trade off is controlled by the client, which can define the accepted error in the received answer.

In this paper we propose a method of trading off performance for accuracy using sampling. We suggest a specific way to perform sampling in vertically split datasets and calculate sample size given acceptable error. 
Performance improvements of the method are shown on a calculation of intersection set cardinality, together with a proposed heuristic algorithm. We also discuss a case of non-cooperative data providers and adaptation of the proposed algorithm for differential privacy calculations.


\noindent {\bf Our goal}: given a set of $k$ data providers $$DP=\{D_1, D_2, \ldots, D_k\},$$ which data records share a common identifier $w$ and a set of predicates $P=\{p_1, p_2, \ldots, p_k\}$, find the size of intersection $\cap_i p_i(D_i)$ where $p_i(D_i)$ is a set of all records in $D_i$ that satisfy predicate $p_i$. 

We assume that the number of records that are not present in all datasets is very small compared to the total number of records and thus, 
for the sake of simplicity, the data providers in $DP$ are assumed to contain the same records with size denoted by $N = |D|$, i.e. vertical split of the data.

The size of the intersection might not be calculated precisely, but up to a given accuracy $\epsilon$ provided by the client querying the data providers. 
For example, the client that performs a query is interested in an integer percentage of people from the entire population having both criminal record and specific genetic mutation, i.e. the results can be rounded to a closest percentage. 
Intuitively, this relaxed requirement should result in more computationally and network efficient algorithms. The optimization of computation time and network traffic are important, as mentioned above, in the commonly used public-cloud deployment; the user is charged for computation time (also called CPU time) and network traffic. Thus, minimization of those parameters is a very attractive algorithm property and present a trade-off with accuracy of the intersection set size.

\noindent {\bf Related work and our contribution:} 
Conjunctive queries over distributed data\-bases were extensively researched (see as an example~\cite{Koutris11parallelevaluation, leskovec2014mining, Dean04mapreduce:simplified, DBLP:journals/corr/AmelootGKNS14, Woodruff_whendistributed}). Previous research has references both the aggregation techniques and performance issues of such queries. 
However, those works are based on an assumption that data\-bases freely share the data, which does not hold in a settings of non-cooperative data providers.

There were a number of works describing privacy preserving calculations of intersection size, see~\cite{Narayan12djoin:differentially, Freedman04efficientprivate, Kissner05privacy-preservingset, Clifton_secureset, 6517175, Sepehri12102014, Agrawal:2003:ISA:872757.872771}. Most of those works described a protocol that performs an exact calculation of the intersection set using Secure Multiparty Computations (MPC), first investigated by~\cite{4568207} and later generalized to multiparty computation. Performing MPC protocols allows efficient calculation of the intersection size, however, when used by itself, it is also posssible to leak information about specific items in the datasets. For instance, the user can issue a query knowing with a single possible answer and check whether the intersection set is empty or not. Such technique allow the user to identify the presence or absence of a specific item in the datasets. 
A technique that can be used to hide a presence of individual items in the dataset is differential privacy (\cite{DBLP:conf/tcc/DworkMNS06}). 
 
While some of the above works (\cite{Narayan12djoin:differentially}) performed inexact calculations of the intersection set, the source of approximation was to preserve differential privacy of the results. For instance, \cite{Freedman04efficientprivate} applied Oblivious Transfer (OT) protocol to approximate the size of the intersection of two data\-bases, where OT has lead to the inexact result. Our approach is to utilize the relaxed requirement of providing inexact result to improve the computational complexity of queries.

Sampling as a way to cope with huge volumes of data or to increase computation performance were considered in a number of different contexts.
(\cite{Gibbons_distinctsampling, Gibbons97fastincremental, Poosala97selectivityestimation}) and Section 4.2 of~\cite{leskovec2014mining} discussed sampling as a means to cope with massive datasets by creation of a histogram-based estimators. Once the estimator is created, when possible, the data\-base performed estimation calculations on the histogram. 
Reservoir stream sampling algorithms (\cite{Vitter:1985:RSR:3147.3165, 1701656, 10.2307/2346966}) were developed to provide a sample of online data. The techniques fill in the sample (``reservoir'') by knocking out existing sample items from the sample with reducing probability as the sample begins to fill. 
(\cite{Broder97onthe, Broder98filteringnear-duplicate}) initiated usage of document sampling for document similarity comparison on Internet scale. Ideas were later developed into min-wise sampling techniques (\cite{DBLP:conf/pods/PaghSW14, Cohen97size-estimationframework, Cohen906coordinatedweighted, DBLP:journals/corr/abs-1206-5637}), which appear to be more suited for horizontal split datasets. 
In addition, a considerable corpus of work exists on concentration inequalities in scope of Machine Learning. Concentration inequalities investigate the relation between the size of the sample and its statistical similarity to the entire dataset. PAC-Bayes bounds on hypothesis error as a function of sample size were provided in~\cite{McAllester_PACBayes, McAllester99pac-bayesianmodel, Langford02pac-bayes}. 

In this paper, we consider the case of approximate calculations in distributed data\-bases where the data is split vertically. We suggest an algorithm that takes advantage of a lack of accuracy in a distributed answer to considerably speed up the queries. 
We show a use-case of an algorithm on intersection set cardinality calculations both with and without privacy considerations.
The algorithm adopts sampling techniques from (\cite{Gibbons_distinctsampling, Gibbons97fastincremental}) and uses techniques similar to PAC-Bayes bounds from (\cite{McAllester_PACBayes, McAllester99pac-bayesianmodel, Langford02pac-bayes}) to decide on a mimimal, representative sample size.
Performance improvement might be especially significant in privacy preserving setting, where the calculations take considerable time.
Specifically our contributions are:
\begin{itemize}
	\item Suggested efficient method for approximate, distributed calculations with vertical-split datasets.
	\item Proposed a method of choosing sampling size given a required error level and provided simulation results of comparing the error level of different sample sizes.
	\item Suggested a heuristic algorithm of speeding up the intersection set size calculation based on bounds convergence and showed  its adoptation for privacy preserving intersection set size calculation.
\end{itemize}

The rest of the article is structured in the following way. Section~\ref{sec:Naive Algorithm} defines a na\"ive algorithm for calculation of exact intersection set size, Section~\ref{sec:sampling} relaxes the requirement to calculate the exact size of intersection set and defines theoretical bounds for {\it inexact} intersection. Section~\ref{sec:experiments} presents results of experiments that validate the proposed sampling method. Section~\ref{sec:bounded estimation} describes a heuristic algorithm for calculation of intersection based on the bounds of intersection size. Finally, Section~\ref{sec:privacy} discusses privacy issues in the described algorithms. The paper is concluded in Section~\ref{sec:concluding remarks}.

\section{Na\"ive Algorithm}
\label{sec:Naive Algorithm}
The idea of the Na\"ive Algorithm is simple: iterate over data providers $DP=\{D_1, D_2, \ldots, D_k\}$ and exchange a mo\-no\-to\-nous\-ly decreasing set of keys that satisfy all predicates up to the current iteration.
The data provider then checks which of the received keys answer the corresponding predicate and returns a new list to the client. The client continues the process until all data providers were queried and the intersection set is found. 
In this way, the client calculates the intersection set iteratively. Denote this set by $L_i$ where $i$ is the iteration number. In other words, $L_1 = p_1(D_1)$ where $p_i(D_i)$ is a set of records in $D_i$ that satisfy predicate $p_i$, and $L_i = L_{i-1} \cap p_i(D_i)$.

The Na\"ive Algorithm can also be performed in a parallel way, where the client receives a set of records that satisfy the corresponding predicate from each data provider. The client then calculates the intersection set. 
Comparing the sequential and parallel na\"ive algorithms, sequential algorithm optimizes the network load and also improves CPU time, but not a wall-clock time of the algorithm (i.e., the time between the beginning of the operation and the time the client gets the final answer). 
Unless stated otherwise, in the rest of the section we consider only the sequential version of the algorithm.

Despite its virtue of being simple, the na\"ive algorithm has a few drawbacks.
\begin{itemize}
	\item The amount of information transferred between the client and the data providers is relatively large, as the entire set of keys in the intersection set is sent.
	\item The first data providers evaluate the given predicate over their entire dataset, which is expensive. 
	\item The wall-clock time of the sequential algorithm is linear in a number of data pro\-vid\-ers and the size of the data.
	\item The algorithm completely exposes information both between data providers and between data providers to the client.
\end{itemize}
As mentioned above, those drawbacks are even more extreme in a current common practice, where a public cloud infrastructure is used for calculations, as in a public cloud infrastructure the client is charged for computation time (CPU time) and for network traffic. Thus, there is a monetary incentive to minimize those two values.



Both the sequential and the parallel na\"ive algorithms described above calculate the intersection set exactly, thus having a relatively high network traffic and CPU time demands. The next section discusses a way to reduce computation and network complexity by calculation of an estimation of the intersection size.

\section{Reducing Data Size by Sampling}
\label{sec:sampling}
Relaxing the requirement for exact calculation of the intersection, this section discusses a way to calculate a smaller intersection set whose size can be extrapolated to the intersection size of the entire population.
Instead of calculating the intersection of the entire datasets, perform a sample and calculate the intersection of the sample. Then, scale up the size of the sample intersection to the size of the intersection for the entire population. The results of scale up will be an estimate for the intersection size. 

The requirements from the sample are clear: the sample should be as small as possible while truthfully representing the entire population (given the definition and the error of representation). Those two requirements present a trade-off between sample size and the accuracy of algorithm results.
A number of different sampling techniques were developed over the years: on-line (reservoir) stream sampling algorithms (\cite{Vitter:1985:RSR:3147.3165, 1701656, 10.2307/2346966}), Histogram-based estimators (\cite{Gibbons97fastincremental, Gibbons_distinctsampling, Poosala97selectivityestimation}) and min-wise sampling techniques (\cite{Broder97onthe, Broder98filteringnear-duplicate, DBLP:conf/pods/PaghSW14, Cohen97size-estimationframework, Cohen906coordinatedweighted, DBLP:journals/corr/abs-1206-5637}) which are more suited for horizontal split datasets.
In our settings, it seems natural to choose hash function based technique, which in a sense a randomized version of histogram.

To provide sufficient improvements in performance, the size of the sample (denoted by $m$) should be much smaller than the average size of the datasets ($\mid \hat DP \mid$): $m << \mid \hat DP \mid$. Therefore, if each data provider performs its own sampling, the overlap between the samples will be very small. In other words, the probability of the data record (person in our example) appearing in the sample is low. For every data provider $i$
	\[
		\forall x \in D_i : P( x \in S_i) = \frac{m}{\mid \hat DP \mid},
	\]
	where $S_i$ is a sample from the dataset of the data provider $i$. Leading to the probability of the item appearing in all samples being (assuming that all data providers contain the same records): 
	\[
		 P( x \in \cap_{j=i}^k S_i) = \prod_{l=1}^k \frac{m_l}{\mid D_l \mid},
	\]
	where $n_l$ and $\mid D_l \mid$ are a sample size and a dataset size of a data provider $l$. Assuming, for simplicity, that the sample size and dataset size are equal for all data providers, the probability becomes:
	\[
		P( x \in \cap_{i=1}^k S_i) = \left( {\frac{m}{\mid \hat DP \mid}} \right)^k.
	\]
	As this probability is low, the size of the intersection will be essentially zero for most of the samples. The solution is to make all data providers create the same sample. In order to do that, we will use a sampling technique based on hashing (\cite{Gibbons_distinctsampling}). 
 The technique works in the following way:
\begin{enumerate}
	\item The client defines an accuracy threshold for the calculation. 
	\item Using techniques described below, the size of the sample is determined: $m$. 
	\item The number of hash buckets is $b = \left\lceil \frac{\mid \hat DP \mid}{m} \right\rceil$.
	\item Pick a hash function $H$ that will distribute datasets of data providers $D_i$ into $b$ buckets. $H$ hashes only the ID of the records and not the entire data, as the ID is the shared information across different data providers.
	\item The client sends each data provider 3 parameters: $H$, the number of the bucket that was chosen randomly and predicate $p_i$.\footnote{It is possible to always use a predefined bucket with a downside that for a given size, there is a single representative sample.} This will ensure that all data pro\-vi\-ders use the same sample.
	\item Each data provider evaluates its predicate on the re\-cords in a given bucket ($p_i(D_i)$) and sends the results to the client.
	\item The client performs the intersection calculation on the received results.
\end{enumerate}

As client acceptable error defines sample size, which is the bucket size, it is possible to pre-calculate a number of buckets for different error value. 
This will eliminate the need for sequential scan of the entire database on each query. However, it will also mean that the client will get approximately the required error.

Notice that just like the Na\"ive algorithm sampling can be done both in parallel and sequential ways. Following the similar idea of the Na\"ive algorithm, in the parallel version of sampling algorithm each data provider evaluates the predicate over the entire given bucket, where in the sequential version, the client sends the current intersection set (which is a subset of the given bucket) to the next data provider, which evaluates the given predicate only over the received subset.
However, in contrast to the Na\"ive algorithm, where the sequential variant significantly reduces the network load and CPU time, in sampling algorithm the reduction is much smaller, as the core underlying idea of the sampling algorithm is minimization of the amount of records participating in the intersection calculation.
Importantly, the accuracy of the intersection size estimation is the same in both sequential and parallel versions of the sampling algorithm.

Sampling improves both the computation time and network traffic over the Na\"ive algorithm by the factor of $\frac{m}{\mid \hat DP \mid}$. 
Thus, one important question still remains unanswered: what should be the size of the sample given an accuracy threshold?
The following sections define bounds on the sample size such that with high probability it will represent behavior similar to the entire dataset. Two different cases are considered: sample size is small compared to the dataset size (Section~\ref{subsec:small sample}) and sample size is comparable to dataset size (Section~\ref{subsec:big sample}). 
Notice that in both cases the sampling is performed in the same way as explained above. However the bound used to calculate the sample size is different for those two cases.

\subsection{Selecting a Sample Size}
\label{subsec:selecting a sample size}
	Selection of the sample size is driven by the trade-off between the accuracy of the sample and the performance of calculations. The bigger the sample, the more accurate it is, but also the larger the computation time required to calculate the intersection. Using a bound on the error as a function of the sample size, it is possible, given a limit on an acceptable error, to choose the size of the sample set.

\subsubsection{Bound for Small Sample from Large Dataset}
	\label{subsec:small sample}
	When a sample size is very small compared to the dataset size, the sampling can be approximated by independent sampling of the same size with replacement. The approximation can be done, as the probability of any record being a part of the sample goes from $\frac{1}{N}$ for the first pick to $\frac{1}{N-m}$ for the last pick. If $N >> m$ then $N-m \approx N$, and therefore, the probability is approximately ($\frac{1}{N}$), which is the same probability for a record to be picked when sampling with replacement. 
	The reason to perform such approximation is due to a fact that it is simpler to provide a bound for independent samples with replacement rather than for sampling without replacement.

	The size of the sample should be big enough that the sample will be a good representative of the entire dataset for the intersection calculations. How can the size be estimated?

	Let us consider a sampling from the dataset $D_i$ according to the uniform distribution. Notice that even though the sampling is done according to uniform distribution, as described in the sampling algorithm (Section~\ref{sec:sampling}), the datasets themselves might be drawn from other distributions. The provided bounds still hold, as the sample ``similarity'' to the entire dataset depends only on its size. If a different, non-uniform sampling technique is used, the bound might be changed to use more general Bernstein inequalities (\cite{Bernstein}).

	Define $X_i$ to be a random variable that the sampled item $w_i$ satisfies condition $L$. Then, the average value of the sample, denoted by $\hat{X}$, should be close to the mean value, $\mu$, of the entire dataset. 
	Notice that in the binary case, the average is simply the number of data records that satisfy a given predicate divided by the data size.
	In order to bound the difference between the average value of the sample and the mean value of the dataset, we can use concentration inequalities. Namely using results by Hoeffding (\cite{Hoeffding62probabilityinequalities}) if $X_1, X_2, \ldots, X_m$ are independent and $i=1, 2, \ldots, m : 0 \leq X_i \leq 1 $, then:
	\begin{equation}
		Pr( \bar{X} - \mu \geq \epsilon) \leq { \left(\frac{\mu}{\mu +\epsilon}\right)^{(\mu + \epsilon)} \left(\frac{1-\mu}{1-\mu-\epsilon}\right)^{(1-\mu-\epsilon)} } ^ n  \leq e^{-2 m \epsilon^2}.
		\label{eq:full hoeffding}
	\end{equation}

	For simplicity, for now we will consider only the right-side of the equation, i.e.
	\begin{equation}
		Pr( \bar{X} - \mu \geq \epsilon) \leq e^{-2 m \epsilon^2},
		\label{eq:short hoeffding}
	\end{equation}
	where $m$ is the size of the sample and $\epsilon$ measures the ``resemblance'' of the sample to the mean of the dataset. Notice that the equation does not depend on the size of the dataset, as it is considered much larger than the sample. The bound in Equation~\ref{eq:short hoeffding} is used by the client when it issues queries to the data providers by defining both the acceptable error ($\epsilon$) and the target probability of exceeding the acceptable error.

\subsubsection{Sample Size is Comparable to a Dataset Size}
	\label{subsec:big sample}
	In cases where sampling size is comparable to the dataset size, it is possible to develop bounds directly for sampling without replacement. In the rest of the section, we show two bounds on sampling without replacement, one is based on a reduction of ``randomness'' of the data and the second one is based on a direct counting technique. 
	Those bounds can be expected to be tighter than those based on reduction to independence or bounds for sampling with replacement. 
	The reason for this as follows. Assume that $k$ points were sampled out of $N$ points without replacement. The next point is to be sampled from a set of $N-k$ rather than $N$ points, which would be the case in sampling with replacement. The successive reduction in the size of the sampled set reduces the ``randomness'' of the newly sampled point as compared to the independent case, and also introduces dependency between samples. 
	Whereas, the bound provided in Equation~\ref{eq:short hoeffding} does not depend on the dataset size and does not take the reduction in population size into consideration. 
	This intuition is at the heart of Serfling's improved bound (\cite{serfling1974}) which is stated next. The result holds for general bounded loss functions and is established by a careful utilization of martingale techniques combined with Chernoff's bounding method.

	\begin{theorem}
		\label{theorem:Serfling}
		(Serfling) Let $C = {c_1, c_2, \ldots, c_N}$ be a finite set of non-negative bounded real numbers, $|c_i| \leq B$. Let $Z_1, Z_2, \ldots, Z_m$ be a random variables obtaining their values by sampling $C$ uniformly at random {\bf without} replacement. 
		Set $Z = (1/m) \sum_{i=1}^{m} Z_i$. Then,
		\begin{equation}
			\label{eq:serfling}
			Pr( Z - {\bf E}Z \geq \epsilon) \leq \exp{\left\{ -\left( \frac{2m\epsilon^2}{B^2} \right) \left( \frac{N}{N-m+1} \right) \right\} }.
		\end{equation}
		Similar bounds hold for $Pr({\bf E}Z - Z \geq \epsilon)$.
	\end{theorem}
	Compared to the bound in Equation~\ref{eq:short hoeffding}, the above bound is always tighter when $N/(N-m+1) > 1$, i.e. when $m > 1$. 

	In our case $c_i$'s are binary variables and the bound could be improved further by using a proof based on a counting argument (\cite{DBLP:journals/jair/DerbekoEM04}).

\subsubsection{Using Bounds to Calculate Sample Size}
\label{subsec:using bounds to calculate sample size}
\begin{figure}[h]
		\centering
		\includegraphics[width=0.5\textwidth]{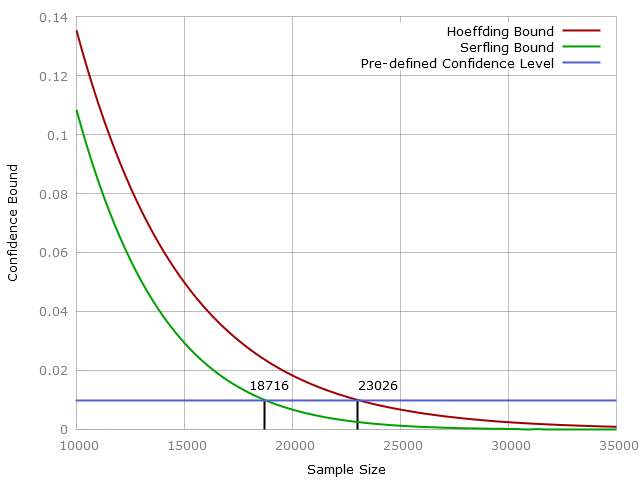}
		\caption{A comparison of Hoeffding and Serfling bounds, where $\epsilon = 0.01$, $N = 100000$.
		The error value is preset and then the sample size that fits this value is picked. In tighter bounds, the sample size will be smaller.}
		\label{fig:bounds comparison}
	\end{figure}

	Figure~\ref{fig:bounds comparison} presents a single example of comparison between the above bounds (\ref{eq:short hoeffding} and \ref{eq:serfling}). As expected, the Serfling bound is tighter than the Hoeffding bound and thus, using this bound for calculation of sample size will result in a smaller sample set. 

	Now we can briefly describe the process of using those bounds. 
	We assume that the client attempts to minimize the size of the sampling sets, as this also minimizes the network load and CPU time, both of which are chargeable in public cloud environments. 
	The client defines a value of an acceptable error for the combined distributed query ($E$) and a ``confidence'' of the error.\footnote{In a very similar way to PAC bounds~(\cite{Valiant84atheory}) in Machine Learning.}
	The bounds defined in (\ref{eq:short hoeffding}) or (\ref{eq:serfling}) provide a ``confidence'' that the error will be smaller than $\epsilon$ given a sample size. 
	Now the client uses the bound from (\ref{eq:short hoeffding}) or (\ref{eq:serfling}) to determine the minimum sampling size that allow the required error with ``high enough'' confidence. 
	Using a predefined confidence and a chosen bound, the client can find the minimum value of sample size that will provide the required error with the required confidence. For example, consider Figure~\ref{fig:bounds comparison}, where for dataset size of $100,000$ the client choose $\epsilon=0.01$ and confidence of $0.01$. Thus, using the simplified Hoeffding bound (\ref{eq:short hoeffding}) the minimal sample size is $23,026$ and using the Serfling bound (\ref{eq:serfling}) the minimal sample size is $18716$.\footnote{This example also shows the advantage of tigher bounds, which in this case is $23\%$ of a sample size.}

	After the intersection of sample sets is calculated, there is a need to estimate the size of the intersection of the entire datasets.
	As mentioned above, the mean of the binary random variable is also the number of records that belong to a set defined by the predicate divided by the size of the sample. Thus, the ratio of the records in a sample that satisfy a given predicate is close to the ratio of the records that satisfy the predicate in the entire dataset of a given data provider.
	This leads to the conclusion that the ratio of the sample intersection set size to the sample size should be the same as the ratio of the intersection set size to the size of entire dataset. 
	Let $\hat{S_i}$ be a sample of $D_i$ of size $m$, then
	\begin{equation}
		\label{eq:intersection size}
		|\cap_{i=1}^{k}{D_i}| = |\cap_{i=1}^{k}{ \hat{S_i} }| \frac{|\hat{DP}|}{m}.
	\end{equation}

	Notice that even though the absolute error in the intersection estimation of the entire dataset is proportionally larger than the error in estimation of a sample, the relative error will remain the same.
	The reason for the error to remain the same in a process of intersection calculation is due to the sampling method. Even though the sample from each data provider has the same (potential) relative error, the error is unique for the sample and thus, it does not accumulates when the intersection is calculated.

	As $\epsilon$ in the above bounds (Equation~\ref{eq:short hoeffding} and~\ref{eq:serfling}) is a bound on difference in estimation, it is a relative error. Therefore, if the client defines an acceptable error ($E$) as an absolute error, it can be easily translated: $\epsilon = \frac{E}{\hat{D}}$. 

\section{Experiments}
\label{sec:experiments}
	We have performed a number of experiments to show the utility and error resulted from sampling.
	The experiments were performed on simulated data and validated on the Adult dataset (\cite{Lichman:2013}). 
	The described sampling technique is targeted at large datasets with many records, such that calculation of predicates over the entire dataset is wasteful. For such datasets, it is much more practical to test the algorithm on simulated data, which can be generated on any required scale.

	\subsection{Experiments on Simulated Data}
	\label{subsec:experiments on simulated data}
	For experiments on simulated data, a number of datasets were generated with random values of the predicates and then the intersection size was calculated according to each algorithm. The methodology works as follows.
	\begin{itemize}
	\item Generate a dataset of a given size. All data providers assumed to share the same set of records with possibly a different data per record.
	\item For each data provider generate a set of predicate values randomly, such that the frequency of ``true'' values is as defined.
	\item Calculate the intersection iteratively, i.e. by addition of a single datasets at a time. This is a both a simpler way of implementation (as opposite to parallel calculation) and also allows observation of convergence rate of the intersection size.
	\end{itemize}
	All experiments were averaged over 10 runs, standard deviation was calculated and drawn on the resulting graphs.

\begin{figure}[h]
		\centering
		\includegraphics[width=0.5\textwidth]{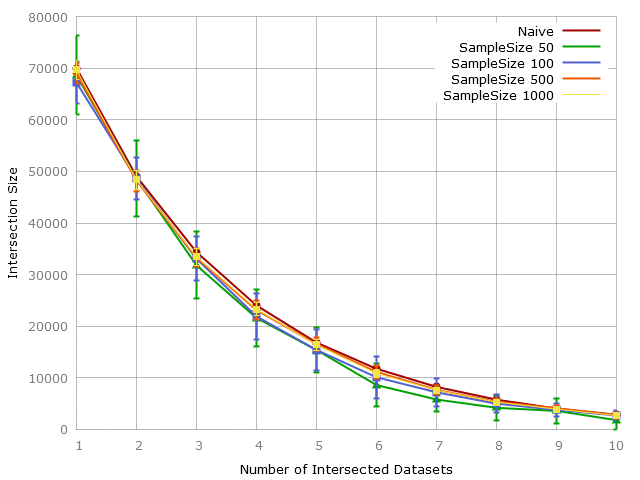}
		\caption{A comparison of the Na\"ive and the Sampling algorithms. Dataset size was set to $N=100,000$, number of datasets is $k=10$, ratio of predicate satisfaction in each dataset is $0.7$. The Y axis shows the size the intersection estimation. The X axis show the number of the data sets that participated in the intersection. Error bars show standard deviation over 10 different executions.}
		\label{fig:algorithms comparison}
	\end{figure}

	Figure~\ref{fig:algorithms comparison} shows the intersection size as calculated by the sequential na\"ive algorithm and estimation by sampling of various sizes. The graph provides a high-level, visual practical validity of the approach. Even a relatively small sample sizes ($1\%$ of the dataset size) estimate intersection size close to the real value.

\begin{figure}[h]
		\centering
		\includegraphics[width=0.5\textwidth]{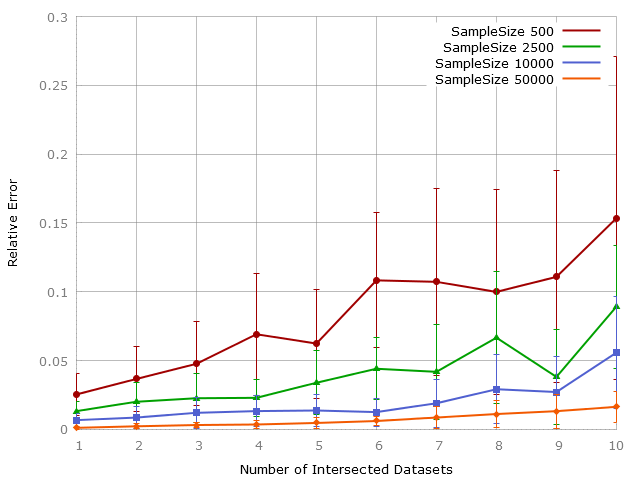}
		\caption{A comparison of errors of different sample sizes. Dataset size was set to $N=100,000$, number of datasets is $k=10$, ratio of predicate satisfaction in each dataset is $0.7$. Error bars show standard deviation over 10 different executions.}
		\label{fig:algorithms error comparison}
	\end{figure}

\begin{figure}[h]
		\centering
		\includegraphics[width=0.5\textwidth]{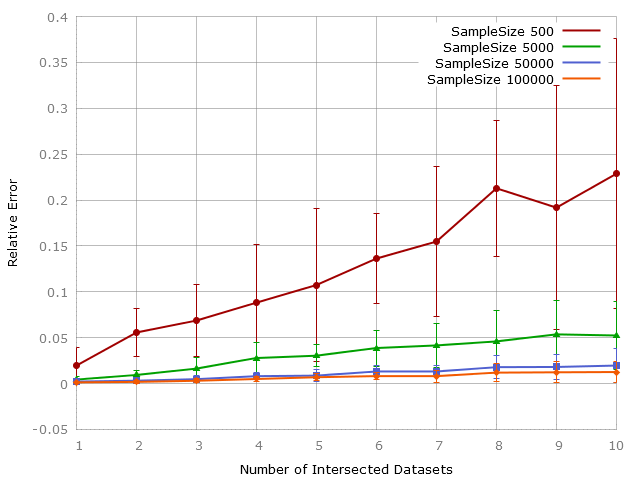}
		\caption{A comparison of errors of different sample sizes. Dataset size was set to $N=1,000,000$, number of datasets is $k=10$, ratio of predicate satisfaction in each dataset is $0.7$. Error bars show standard deviation over 10 different executions.}
		\label{fig:algorithms error comparison 1M}
	\end{figure}

\begin{figure}[h]
		\centering
		\includegraphics[width=0.5\textwidth]{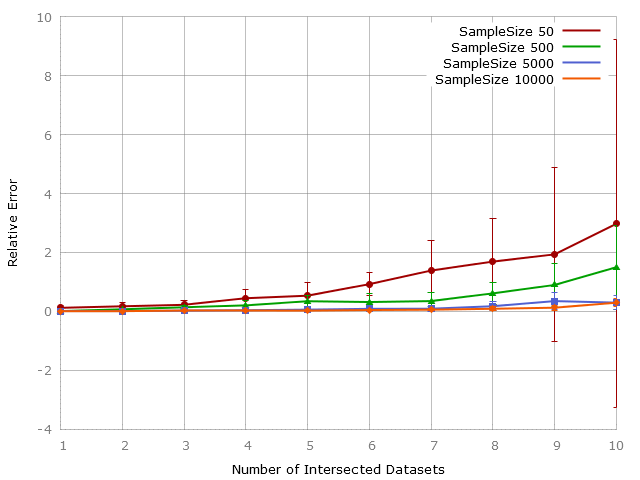}
		\caption{A comparison of errors of different sample sizes. Dataset size was set to $N=100,000$, number of datasets is $k=10$, ratio of predicate satisfaction in each dataset is $0.5$. Error bars show standard deviation over 10 different executions.}
		\label{fig:algorithms error comparison 1M R05}
	\end{figure}

\begin{figure}[h]
		\centering
		\includegraphics[width=0.5\textwidth]{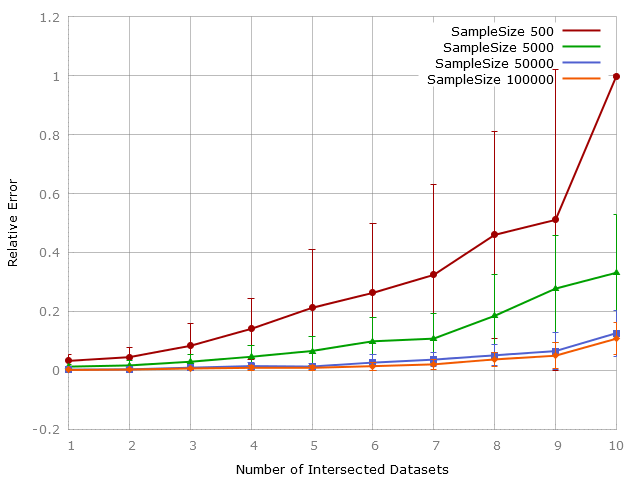}
		\caption{A comparison of errors of different sample sizes. Dataset size was set to $N=1,000,000$, number of datasets is $k=10$, ratio of predicate satisfaction in each dataset is $0.5$. Error bars show standard deviation over 10 different executions.}
		\label{fig:algorithms error comparison 100K R05}
	\end{figure}

	In order to focus on the error caused by sampling and show the difference in sample size more clearly, we have performed a number of experiments showing relative error between estimation and the real intersection size.
	The Na\"ive algorithm calculated was taken as a baseline for error calculations. The Sampling algorithm was executed with a number of sample sizes, each one showing the relative error from the Na\"ive algorithm results. 
	Figures~\ref{fig:algorithms error comparison}, \ref{fig:algorithms error comparison 1M}, \ref{fig:algorithms error comparison 100K R05}, and~\ref{fig:algorithms error comparison 1M R05} show the relative error of the sampling algorithm with various sample sizes. The Y axis shows the error of the sampling, i.e. |sampled estimation - intersection size|  / (intersection size),\footnote{The error is relative but not normalized. It is possible to multiple the error by 100 to translate it into percentages from the intersection size. Notice that as the intersection might be relatively small, the error in percentage might be large. Comparing it to the dataset size will result in a much smaller error values.} while the X axis is the number of datasets in the intersection. The graphs compare datasets sizes of 100,000 (Figures~\ref{fig:algorithms error comparison} and \ref{fig:algorithms error comparison 100K R05}) and 1,000,000 (Figures~\ref{fig:algorithms error comparison 1M} and \ref{fig:algorithms error comparison 1M R05}) records with predicate satisfaction frequency of $0.7$ and $0.5$, i.e. in each data provider, $70\%$ or $50\%$ of the records satisfy the corresponding predicate. As mentioned above, each graph is an average of 10 different runs and standard deviation is depicted by error bars on the graph.

	The graphs show that, as expected, larger samples result both in smaller error and smaller standard deviation. However, it also can be seen, that the error quickly converges as a function of sample size. In some cases, even for sample of $1\%$ from entire dataset, it is possible to achieve reasonable error.  
	Overall, sampling $1/10$ or $1/5$ of the dataset resulted in errors of approximately $0.01$ from the intersection size, which is an order of a single percent. Thus, queries of a type: ``What is the approximate percentage of people ... ?'' can be answered using only tenth of a data.

	\subsection{Experiment on a Real Dataset}
	\label{subsec: experiment on real dataset}
	In addition to experiments on simulated datasets, we have tested the methodology on Adult dataset from UCI Machine Learning Repository (\cite{Lichman:2013}). The dataset contains data from census bureau, where each record has data about different person. This fits a description of the setup we have described.
	The dataset contains 32,562 instances. As a test case we have used an intersection of the following predicates: age $\ge$ 30, marital status: Never-married, Sex: Female or Male in two different tests and income $>$ 50K. The exact intersection set size is 252 for males and 139 for females. Since the dataset was not used for classification task, it allowed us to use income as one of the predicates. In addition, the intersection was of 4 predicates, as addition of more predicates resulted in a small or empty intersection set due to a limited dataset size.
	
\begin{figure}[h]
		\centering
		\includegraphics[width=0.5\textwidth]{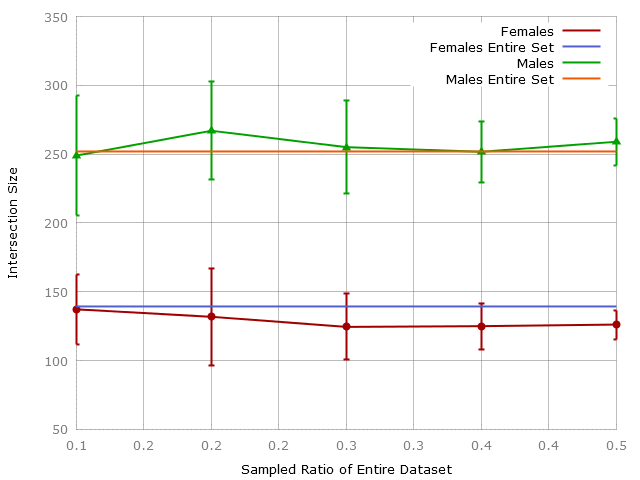}
		\caption{Comparison of intersection set size to the estimated size for different sample ratio. The experiments were performed 10 times with different samples. Average value is shown on the graph together with standard deviation results.}
		\label{fig:adult sampling}
	\end{figure}

	Samples of different sizes were drawn with sample size ratio going from 0.1 to 0.5. The intersection set size was calculated from the drawn sample.
	Figure~\ref{fig:adult sampling} depicts the results of the testing. While the accuracy of the estimation does not necessary improves by taking larger samples, the standard deviation becomes smaller. The accuracy improvement is most probably caused by the small size of the intersection, while improvement in standard deviation fits the results shown in simulated datasets. Overall, the accuracy of the intersection set size calculation verify the validity of the approach.

\section{Heuristic Algorithm for Bounded Estimation}
\label{sec:bounded estimation}
Previous sections described exact and approximate ways of calculating the size of the intersection. This section pre\-sents a heuristic algorithm that attempts to optimize the calculation of the intersection by not performing the calculations for all datasets. The simplest case of heuristic is when the intersection is calculated iteratively and at some point the intersection set is empty. Clearly, the calculation can be stopped at this stage. Following is a heuristic algorithm for the intersection calculation that attempts to stop at the earliest possible point.

The size of the intersection depends on the sizes of the sets from each data provider that answer the corresponding predicate and on the correlation between those sizes. Below we suggest a heuristic algorithm for estimation of the intersection set size. The algorithm starts with an accuracy parameter, the ratio of records that satisfy the relevant predicate in each data set, and the lower and the upper bounds on the intersection size. The algorithm iteratively tightens the bounds until the difference between the bounds is smaller or equal to the accuracy parameter. Then the algorithm stops and returns the middle value of the range between the lower and the upper bound. 

\subsection{Upper and Lower Bounds on the Intersection Size}
\label{sec:bounds}
Let $p_i$ be the predicate that is associated with a data provider $D_i$. Then, $p_i (D_i)$ is the set of members in $D_i$ that satisfy $p_i$. Also, $\hat{p_i}$ will denote the fraction of the members that satisfy $p_i$ in $D_i$, i.e. $\hat{p_i} = \frac{\mid p_i(D_i) \mid}{\mid D_i \mid}$. Now, let us define the bounds on the size of the intersection set $J$ for the sake of the algorithm iterations. 
For simplicity, we assume that data provider datasets $D_1, D_2, \ldots, D_k$ contain {\em exactly} the same records. The bounds will hold when the difference in members between datasets is small.

As the intersection size is at most the size of the smallest predicate set, the size of the intersection is bounded from above by the following bound:
\begin{equation}
	\label{eq:upper bound}
	\mid J \mid \leq min_i \mid p_i (D_i) \mid.
\end{equation}

The lower bound on the intersection of two sets $D_1$ and $D_2$ is (\cite{focsBook1994})
\[
	\mid D_1 \cap D_2 \mid \leq 
	\begin{cases}
		(\hat{p_1} + \hat{p_2} - 1) \frac{\mid D_1 \mid + \mid D_2 \mid}{2}, & \text{if $\hat{p_1} + \hat{p_2} \ge 1$}.\\
		0, & \text{otherwise}.
	\end{cases}
\]
In case of 3 sets, the lower bound becomes:
\[
	(\hat{p_1} + \hat{p_2} - 1) + \hat{p_3} - 1,
\]
which in general is
\begin{equation}
	\label{eq:lower bound}
	\mid \cap_{i=1}^{k}{D_i} \mid \geq max \left\{ \left(\sum_{i=1}^k \hat{p_i} - (k - 1)\right) \mid D \mid, 0 \right\},
\end{equation}
where $\mid D \mid$ is the average size of the datasets (as they contain the same set of records). 


\subsection{The Heuristic Algorithm}
\label{sec:heuristic algorithm}
As described above, the iterative step of the algorithm is to tighten the bounds on the intersection size. The algorithm will then stop once the difference between bounds is smaller than the required accuracy. 
As the upper bound (Equation~\ref{eq:upper bound}) is the minimal predicate set, an iteration step will attempt to make the minimum value smaller by calculating the intersection between the two smallest predicate sets. The intersection of two sets will be smaller or of the same size as minimal predicate set and will replace those two sets in the bound.

Calculating the intersection of two minimal predicate sets is also a good technique for increasing the lower bound (Equation~\ref{eq:lower bound}) as well.\footnote{This can be seen especially in cases where there are two or more predicate sets that are smaller than $0.5$. In this case the lower bound of their intersection is 0.}
Notice that each iteration step of the algorithm also decreases the value of $k$ and thus removes the subtractive member of the lower bound. 

The algorithm steps are as follows.

\begin{enumerate}
	\item Define a required accuracy threshold: $\delta \geq 0.$ As $\delta$ is a relative error, the iterations continue until an absolute error is smaller than $\delta |\hat{DP}|$. When the iterations stop, the middle of the bounds range is returned, therefore, the iterations stop when the distance between the bounds is less or equal to $\frac{\left( \delta |\hat{DP}| \right)}{2}.$ 
	\item Calculate $p_i(D_i)$ for every data provider $i$.
	\item Calculate the lower ($B_l$) and the upper bounds ($B_u$), given $p_i$.
	\item \label{step:iteration} While $B_u - B_l > \frac{\left( \delta |\hat{DP}| \right)}{2}$ and the number of sets is larger than 1:
		\begin{enumerate}
			\item Let $p_j = min_i p_i (D_i)$ and $p_\ell = min_{i \neq j} p_i (D_i)$. In other words, pick two minimal predicate sets: $j$ and $\ell$.
			\item \label{step:intersect} Calculate the intersection between those two sets.
			\item Replace $p_j (D_j)$ and $p_\ell (D_\ell)$ with the new predicate set: $p_{jl} (D_{jl}) = p_j(D_j) \cap p_\ell(D_\ell)$.
			\item Update new values of $B_u$ and $B_l$.
		\end{enumerate}
	\item If only one set remains $p'$ ($p' = p_{1, 2, \ldots, k}$), then the size of its predicate set, $p' (D') = p' (D_{1, 2, \ldots, l})$, is the size of the intersection set. 
	\item If there is more than one set and $B_u - B_l < \delta |DP|$, then the size of the intersection set to the middle value of $[B_l, B_u]$ range: $\frac{B_u - B_l}{2}$.
\end{enumerate}

The algorithm clearly converges, as the iterations stop when the intersection size is calculated exactly. At this point the lower bound is equal to the upper bound and thus, the accuracy requirement will be satisfied.
Notice that if step~\ref{step:intersect} results in an empty set, the algorithms also stops, as the lower and the upper bounds will be equal and zero.

\textbf{Example:}
	\label{example:algorithm example}
	As an example of the algorithm execution, assume $k=4$, equal size of all data sets $N$ and the required accuracy of $10\%$. 
	Let the respective ratios be $p_1=0.8, p_2=0.9, p_3 = 0.5$ and $p_4 = 0.4$. In this case, the upper bound on the intersection set size will be $N p_4 = 0.4 N$, whereas the lower bound will be $0$, as the sum of all ratios is $2.6$ less than $k-1 = 3$.

	The first iteration of the algorithm will be to find the intersection of $D_4$ and $D_3$. Let us assume that their intersection results in $p_{3,4} = 0.31$. Now, the upper bound of the intersection becomes $0.31 N$, whereas the lower bound increases to $0.8 + 0.9 + 0.31 - (3 -1) = 0.1 N$. The iterations continue until the bounds converge to within $0.1 N$. 
	
\begin{figure}[h]
		\centering
		\includegraphics[width=0.5\textwidth]{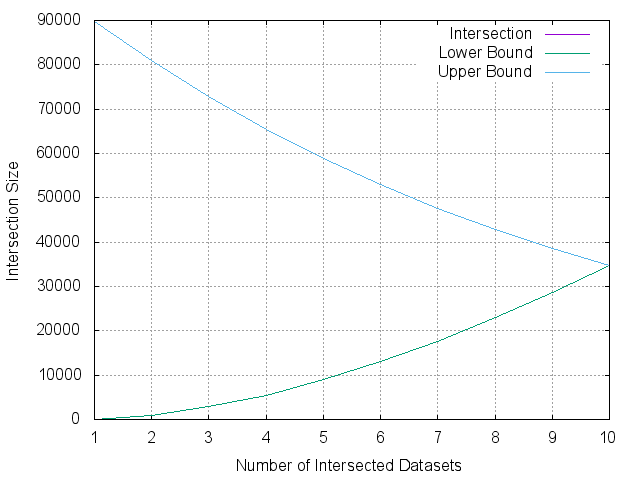}
		\caption{Convergence of the Heuristic Algorithm bounds. Dataset size was set to $N=100,000$, number of datasets is $k=10$, ratio of predicate satisfaction in each dataset is $0.9$ (the high ratio is required for the lower bound to be larger than zero for $k=10$). Notice that in simulated case, the intersection size is co-located with the upper bound of the intersection.}
		\label{fig:heuristic algorithm}
	\end{figure}

	In general, the Heuristic Algorithm will work fine in cases where the ratio of records that satisfy the predicate is high across most data providers, but the intersection size might decrease relatively fast. See Figure~\ref{fig:heuristic algorithm} for example of convergence of the algorithm bounds on simulated data.

\subsection{The Heuristic Algorithm Combined with Sampling }
\label{sec:heuristic algorithm with sampling}
	Clearly, both the computation time and network traffic depend heavily on the calculation of the intersection between the two sets in step~\ref{step:intersect} of the Heuristic Algorithm. To improve the performance of this step it is possible to use the sampling technique from Section~\ref{sec:sampling}. 
	Instead of performing intersection between two sets $i$ and $j$, the algorithm will calculate the intersection of two samples of those sets. However, sampling introduces an error into calculation of the intersection, which has to be related to the accuracy threshold $\delta$ of the heuristic algorithm. 

	Assuming that the error in sampling distributes proportionally among datasets and noting that each iteration ``eliminates'' one dataset, we can define the following changes to the algorithm. 
	Every sample introduces error of $\delta' = \frac{\delta}{k-1}$. To accommodate this error, the algorithm will decrease the required accuracy threshold by this amount on each iteration. This will ensure that required by user accuracy threshold will be honored. 
Two steps are changed in the algorithm:
\begin{enumerate}
	\item Step~\ref{step:iteration} then becomes:\\
		{\em While $B_u - B_l > (\delta - \frac{\delta}{k-1} i) \frac{|DP|}{2},$ where $i$ is the iteration number, and the number of sets is larger than 1.}
	\item Step~\ref{step:intersect} becomes:\\
		{\em Calculate the intersection between those two sets: $j$ and $l$ by sampling with error threshold $\frac{\delta}{k-1}$.}
\end{enumerate}
The rest of the algorithm remains the same, including required accuracy threshold value.	

\section{Privacy of Intersection Calculations}
\label{sec:privacy}
In the described setting there are 2 different types of actors: data providers and client. 
There are different privacy concerns between those actors.
One privacy concern is preserving the privacy of data providers. The client querying the data providers should not learn the identity of the records that are part of the intersection set or the number of records that satisfy a specific predicate in any single data provider. 
Notice that the client should be limited to a reasonable number of queries (sub-linear in a data size), as it is not possible to answer a large number of arbitrary queries while preserving privacy (\cite{Dinur03revealinginformation, Blum05practicalprivacy}).
Another privacy concern is in keeping a privacy of records in a specific data provider from other data providers. For instance, a data provider might be interested in hiding the presence of specific record from other data providers.

First, it is imperative to note that the sampling (see Section~\ref{sec:sampling}) provides a very na\"ive form of privacy. As every data set record has a $1- \frac{m}{N}$ chance of not being included in sample, the presence of the specific record in the data set is not immediately observable. 
Yet, the fact that the client gets record identifiers of each data provider is undesirable. Data providers can encrypt the sampling indices to hide the identity of the records that are sent to the client. Even though an additional encryption will obscure the identity of those items from the client, it still will expose the presence of individual records to other data providers. 
In addition, using an additional knowledge, the client can easily learn the presence of a specific record in the data providers. 

\subsection{Secure Multi-Party Computation (MPC)}
\label{subsec:secure multiparty computation}
One way to preserve privacy is to use secure multiparty computations to calculate the intersection set (\cite{Narayan12djoin:differentially, Freedman04efficientprivate, Kissner05privacy-preservingset, Clifton_secureset, 6517175, Sepehri12102014})
A simple method that uses commutative cryptography (another approach is to use secret sharing) is described in~\cite{secureDistributed2006}. 
Each data provider encrypts record identifiers using its own commutative key and passes the key to other data providers. 
Once all data providers have encrypted all keys, it is possible to find the intersection using encrypted identifiers utilizing commutativity of the encryption and calculate its size.

There are a few downsides to using secure multi party computations. First, performance is heavily impacted by performing secure computations using these methods. 
It is possible to aleviate the performance issues by using the sampling to reduce a number of records in the calculated intersection, as described in Section~\ref{sec:sampling}. 
Thus, the idea is to perform the sampling algorithm and then calculate the intersection in a secure way using the MPC computations between the data providers. 
This method however, will still require calculation of the sampled intersection exactly, without the ability to halt when required accuracy was reached, as it is possible in heuristic algorithm. 
The second drawback is specific for the described use-case where the data is kept in different organizations. Secure multi party computations require direct communication between data providers, which in some cases is very challenging in a real-world deployments due to both security and technological reasons. 
While it is still possible for the client to act as an intermediary between different data providers, such setup also doubles the network traffic. A different approach is only to allow communication between client and data provider with adopted algorithms.

However, the most significant drawback of MPC is that the exact intersection size is calculated. This might allow the client to use additional information to infer a presence of a specific record in the dataset.

\subsection{Differential Privacy}
\label{subsec:differential privacy}
The current de-facto privacy standard is differential privacy (\cite{DBLP:conf/tcc/DworkMNS06}). Informally, the idea of differential privacy is to protect the privacy of individual records in the dataset without any assumption on the additional knowledge. Differential privacy is preserved if it is practically impossible to identify whether the record is present or absent in the dataset from the result of database queries. The most common methods of ensuring differential privacy rely either on addition of a carefully chosen amount of noise to the result or by using an exponential mechanism that chooses the output according to some specific probability distribution.

The protocols that are based on secure multiparty computations perform the exact calculation of the set, which is impossible in differential-privacy settings. This led to several works describing differential-privacy-preserving calculations of the intersection size, see~\cite{Narayan12djoin:differentially, 6517175}. 
Even though those algorithms preserve differential privacy, the requirement to provide an approximate answer in our case allows our scheme to optimize the algorithm for performance by performing secure computations on as little number of the records as possible due to sampling.

The most common method of ensuring differential privacy is an addition of a specifically crafted random noise to the released data (\cite{DBLP:conf/tcc/DworkMNS06, Dwork06differentialprivacy}). 
Moreover, there are two different approaches for the private data release: interactive (\cite{Dinur03revealinginformation})and non-interactive (\cite{Mohammed:2011:DPD:2020408.2020487}). Interactive data release is when the client sends data query to a data provider. The data provider then releases the data to the client while ensuring differential privacy of records in its database. The data provider might decide not to answer a specific queries or to stop answering queries from a given client, as this might impact the privacy. In non-interactive data release, the noise is applied only once and then the data is released to the client. The client can perform any number and any type of queries on the data. 
Intuitively, non-interactive release requires to add more noise to the released data, as it has to be ready for any query, while interactive release can adopt added noise to the results of each query (\cite{DBLP:conf/tcc/DworkMNS06}). 

\subsubsection{Where To Add Noise in Heuristic Algorithm}
\label{subsec:where to add noise}
In the described above sampling algorithm (Section~\ref{sec:sampling}), there are two intuitive locations to add random noise. First, it is possible to add noise in hashing function that assigns records to buckets. Such noise will preserve privacy of individual records, as it will be impossible to distinguish whether a specific record is within the bucket or not present at all in the dataset. A second location to add the noise is during a query processing. 
Since the data provider exposes only the number of records that satisfy a given predicate, the noise can be added to the output after the predicate was evaluated or to the predicate itself [[Ehud: this is not clear, the noise should be fake IDs]]. 

The difference between those two locations is exactly the difference between interactive and non-interactive data release. Adding noise to the hashing function is a one-time operation and has no relationship to the result of the specific predicate. Thus, this noise addition can be seen as non-interactive data release\footnote{It is possible to execute hash function to divide records into buckets on every query. 
However, even in this case, there is no relationship between hash functions and the number of records that answer a specific predicate. Therefore, the added noise is agnostic for the processed query and there is also a performance penalty in constant calculation of hash function over the entire database.}. 
On the other hand, if the noise is added after the predicate is evaluated over records in a bucket, then the noise value can be adapted to the results of predicate evaluation. Thus, the amount of noise can be directly related to the results of a specific query, like in interactive data release settings.

Due to the above reasoning, the next section describes an addition of a noise to the predicate function results.

\subsubsection{Adding Noise to an Output of Predicate Evaluation}
\label{subsec:noisy predicate function}
Differentially private calculation of a counting function (referred in the paper as ``noisy sum''), i.e. counting a number of records that satisfy a given predicate, was considered previously in (\cite{Nissim:2007:SSS:1250790.1250803, Blum05practicalprivacy}). 
(\cite{DBLP:conf/tcc/DworkMNS06}) showed that adding a Laplacian noise $Y \approx Lap(1/\epsilon)$ to the sum query output: $\sum_i x_i + Y$, ensures $\epsilon-$differential private function. Notice that the sensitivity of counting function is $S(f)=1$ and thus, the distribution standard deviation is $ S(f) / \epsilon = 1 /\epsilon$. 

The fast algorithm for privacy-preserving intersection calculation is described below. Heuristic algorithm from Section~\ref{sec:bounded estimation} is used as a basis for privacy preserving algorithm. 

\begin{enumerate}
	\item Use a hash function $H$ to divide data into buckets. In case only a single bucket is used for all queries, only the agreed bucket is kept.
	\item When a predicate is received from the client, evaluate the predicate over the chosen bucket. 
	\item Add random noise according to Laplace distribution to the size of the predicate set. The noisy set-size is shared with the client.
	\item The client gathers results from all data providers and calculate upper and lower bounds. 
	\item \label{step:chosen data providers} The client continues to execute heuristic algorithm and picks two data providers, $j$ and $k$, to perform intersection according to the algorithm step~\ref{step:intersect}.
	\item Perform privacy-preserving, secure intersection between two chosen providers using the below algorithm.	
	\item Update the bounds and continue until the bounds are close enough.
\end{enumerate}

\subsubsection{Differentially Private Set Intersection}
\label{sebsec:differentially private set intersection}
The algorithm used for privacy-preserving, secure intersection set calculation is based on a work of (\cite{Narayan12djoin:differentially}), which allows a differentially private calculation of an intersection over multiple data\-bases.
The algorithm provides computational differential privacy, as it relies on homomorphic encryption and makes some assumptions on computational hardness. This section describes how the algorithm from (\cite{Narayan12djoin:differentially}) can be used in our heuristic algorithm from Section~\ref{sec:bounded estimation} for inexact computations. 

The algorithm describes a basic operation {\em private set-intersection cardinality (PSI-CA)} and then adds noise to ensure differential privacy, thus resulting in BN-PSI-CA. PSI-CA operation is based on usage of homomorphic cryptosystem that allows addition and multiplication by constant (for instance, Paillier's cryptosystem (\cite{Paillier99public-keycryptosystems})). When two data providers, $i$ and $j$, attempt to calculate intersection, $i$ defines a polynomial whose roots are the members of its set, The polynomial coefficients are then encrypted and transfered to $j$ data provider, which calculates $Enc(rP(x_i) + 0^+)$. In other words, it evaluates the polynomial on its set members, multiplies by a constant number $r$ and adds a special string of zeros $0^+$. When transfered back to the first data provider ($i$), it calculates the number of zero string. 
In order to add differential privacy, a random number of dummy values are added to the transfered set by both parties (for details see (\cite{Narayan12djoin:differentially}) rounding the set size to the bucket size in our case. In the end, data provider $i$ learns the noised cardinality and data provider $j$ knows the amount of noise it added to the intersection set size.

Adopting this algorithm for our case, the client performs the following actions. Assume that two data providers are chosen for intersection in step~\ref{step:chosen data providers}: $j$ and $k$. 
\begin{enumerate}
	\item Perform BN-PSI-CA algorithm to calculate a noisy intersection set. Assume that without loss of generality, $j$ will hold the size of a noisy set, where $k$ will hold the amount of added noise.
	\item $j$ sends the size of a noisy set to the client.
\end{enumerate}
In the following iterations of the heuristic algorithm, BN-PSI-CA might be invoked to find intersection size of $j$, $k$ and $r$, and more additional data providers. As due to privacy issues, no single data provider holds the intersection dataset, it is necessary to perform intersection size calculations over again for any additional data provider.

Notice that in cases where data providers do not communicate directly, the client acts as an intermediary for the protocol. In the end, the client has a value of intersection size which is still preserving differential privacy. The client also keeps record that $k$ holds the amount of noise of the last intersection. 
As described in (\cite{Narayan12djoin:differentially}), the noise added in one intersection, might be removed if there is a need to calculate another intersection of $j$ and $k$ with additional data provider.

(\cite{Narayan12djoin:differentially}) performance results show approximately linear dependency of an algorithm run-time in the number of records in the participating datasets. Using results from Section~\ref{sec:experiments}, with a reasonable error, it is possible to reduce the number of records by a factor of 5-10 (according to acceptable error), thus tens of minutes to single minutes. Such a reduction might make a difference between practical and impractical system.

The algorithm accelerate approximate intersection calculation by:
\begin{itemize}
	\item Computationally intensive calculations of intersection sizes are performed on samples. Those calculations are demanding computationally, thus, any reduce in the size of datasets is important. [[Ehud: we should try to find a third-year student to implement this]]
	\item The intersection is not calculate exactly, but rather the algorithm stops when the difference between bounds is close enough.
\end{itemize}

\section{Concluding Remarks}
\label{sec:concluding remarks}
	A steady and rapid increase in the amount of data has resulted in an abundance of various data providers. More data is kept for longer and in more places. In parallel, the awareness of data privacy and security is also on a rise both from government regulation and personal perspective. This trend requires new algorithms and protocols for dealing with distributed data. Those new methods should be both efficient and privacy preserving. 

	The main practical downside of the current privacy preserving computations is the run-time complexity. Both MPC, Oblivious-Transfer and Private-Information-Retrieval methods that are used to perform distributed calculations in privacy-preserving manner significantly increase the running time of the query. While some advances in improving performance were made, the running time still remains a limiting factor. Our proposed method uses the ability to provide an approximate answer to the supplied query and thus to reduce considerably the dataset that is used for computations. 
	This decouples the use of sampling technique for reducing the size of the dataset used for calculations from the protocols designed to preserve privacy of individual records or data providers. As such, most of the above protocols can be used in conjunction with sampling techniques to perform approximate calculations with improved performance. 

	As showed in this work, when acceptable, calculating an approximate answer can significantly improve computation time. In those cases we have suggested a way to use this possibility for approximate answering by using sampling of the datasets. The sampling leads to a dramatic reduction in the size of data required for calculation: $5-10\%$ of the data as observed in simulations. Such reduction of the data size is much more significant when calculations are done in a secure and privacy-preserving way.

\bibliographystyle{abbrv}

\begin{thebibliography}{10}

\bibitem{Agrawal:2003:ISA:872757.872771}
R.~Agrawal, A.~Evfimievski, and R.~Srikant.
\newblock Information sharing across private databases.
\newblock In {\em Proceedings of the 2003 ACM SIGMOD International Conference
  on Management of Data}, SIGMOD '03, pages 86--97, New York, NY, USA, 2003.
  ACM.

\bibitem{focsBook1994}
A.~Aho and J.~D. Ullman.
\newblock {\em Foundations of Computer Science}.
\newblock 1994.

\bibitem{DBLP:journals/corr/AmelootGKNS14}
T.~J. Ameloot, G.~Geck, B.~Ketsman, F.~Neven, and T.~Schwentick.
\newblock Parallel-correctness and transferability for conjunctive queries.
\newblock {\em CoRR}, abs/1412.4030, 2014.

\bibitem{Bernstein}
S.~Bernstein.
\newblock {\em Theory of Probability (Russian)}.
\newblock 1927.

\bibitem{Blum05practicalprivacy}
A.~Blum, C.~Dwork, F.~Mcsherry, and K.~Nissim.
\newblock Practical privacy: The sulq framework.
\newblock In {\em In PODS}, pages 128--138. ACM, 2005.

\bibitem{Broder97onthe}
A.~Z. Broder.
\newblock On the resemblance and containment of documents.
\newblock In {\em In Compression and Complexity of Sequences (SEQUENCES'97)},
  pages 21--29. IEEE Computer Society, 1997.

\bibitem{Broder98filteringnear-duplicate}
A.~Z. Broder.
\newblock Filtering near-duplicate documents.
\newblock In {\em Proc. FUN 98}, 1998.

\bibitem{Clifton_secureset}
C.~Clifton and J.~Vaidya.
\newblock Secure set intersection cardinality with application to association
  rule mining.
\newblock In {\em Accepted for Publication in the Journal of Computer Security,
  IOS}. Press, 2004.

\bibitem{Cohen97size-estimationframework}
E.~Cohen.
\newblock Size-estimation framework with applications to transitive closure and
  reachability.
\newblock {\em Journal of Computer and System Sciences}, pages 441--453, 1997.

\bibitem{DBLP:journals/corr/abs-1206-5637}
E.~Cohen and H.~Kaplan.
\newblock What you can do with coordinated samples.
\newblock {\em CoRR}, abs/1206.5637, 2012.

\bibitem{Cohen906coordinatedweighted}
E.~Cohen, H.~Kaplan, and S.~Sen.
\newblock Coordinated weighted sampling for estimating aggregates over multiple
  weight assignments, 906.

\bibitem{Dean04mapreduce:simplified}
J.~Dean and et~al.
\newblock Mapreduce: Simplified data processing on large clusters, 2004.

\bibitem{DBLP:journals/jair/DerbekoEM04}
P.~Derbeko, R.~El{-}Yaniv, and R.~Meir.
\newblock Explicit learning curves for transduction and application to
  clustering and compression algorithms.
\newblock {\em J. Artif. Intell. Res. {(JAIR)}}, 22:117--142, 2004.

\bibitem{Dinur03revealinginformation}
I.~Dinur and K.~Nissim.
\newblock Revealing information while preserving privacy.
\newblock In {\em In PODS}, pages 202--210. ACM Press, 2003.

\bibitem{Dwork06differentialprivacy}
C.~Dwork.
\newblock Differential privacy.
\newblock In {\em ICALP}, pages 1--12, 2006.

\bibitem{DBLP:conf/tcc/DworkMNS06}
C.~Dwork, F.~McSherry, K.~Nissim, and A.~Smith.
\newblock Calibrating noise to sensitivity in private data analysis.
\newblock In {\em Theory of Cryptography, Third Theory of Cryptography
  Conference, {TCC} 2006, New York, NY, USA, March 4-7, 2006, Proceedings},
  pages 265--284, 2006.

\bibitem{1701656}
P.~S. Efraimidis and P.~G. Spirakis.
\newblock {Weighted random sampling with a reservoir}.
\newblock {\em Information Processing Letters}, 97:181--185, 2006.

\bibitem{Freedman04efficientprivate}
M.~J. Freedman, K.~Nissim, and B.~Pinkas.
\newblock Efficient private matching and set intersection.
\newblock pages 1--19. Springer-Verlag, 2004.

\bibitem{Gibbons_distinctsampling}
P.~B. Gibbons.
\newblock Distinct sampling for highly-accurate answers to distinct values
  queries and event reports.
\newblock In {\em In Proceedings of the 27th International Conference on Very
  Large Data Bases}, pages 541--550.

\bibitem{Gibbons97fastincremental}
P.~B. Gibbons, Y.~Matias, and V.~Poosala.
\newblock Fast incremental maintenance of approximate histograms.
\newblock pages 466--475, 1997.

\bibitem{Hoeffding62probabilityinequalities}
W.~Hoeffding.
\newblock Probability inequalities for sums of bounded random variables, 1962.

\bibitem{secureDistributed2006}
W.~Jiang and C.~Clifton.
\newblock A secure distributed framework for achieving k-anonymity.
\newblock {\em The VLDB Journal}, 15(4):316--333, 2006.

\bibitem{Kissner05privacy-preservingset}
L.~Kissner and D.~Song.
\newblock Privacy-preserving set operations.
\newblock In {\em in Advances in Cryptology - CRYPTO 2005, LNCS}, pages
  241--257. Springer, 2005.

\bibitem{Koutris11parallelevaluation}
P.~Koutris and D.~Suciu.
\newblock Parallel evaluation of conjunctive queries.
\newblock Technical report, 2011.

\bibitem{Langford02pac-bayes}
J.~Langford and J.~Shawe-taylor.
\newblock Pac-bayes {\&} margins.
\newblock In {\em Advances in Neural Information Processing Systems 15}, pages
  439--446. MIT Press, 2002.

\bibitem{leskovec2014mining}
J.~Leskovec, A.~Rajaraman, and J.~D. Ullman.
\newblock {\em Mining of massive datasets}.
\newblock Cambridge University Press, 2014.

\bibitem{Lichman:2013}
M.~Lichman.
\newblock {UCI} machine learning repository, 2013.

\bibitem{McAllester_PACBayes}
D.~McAllester.
\newblock Simplified pac-bayesian margin bounds.
\newblock In {\em Learning Theory and Kernel Machines}, volume 2777 of {\em
  Lecture Notes in Computer Science}, pages 203--215. Springer Berlin
  Heidelberg, 2003.

\bibitem{McAllester99pac-bayesianmodel}
D.~A. McAllester.
\newblock Pac-bayesian model averaging.
\newblock In {\em In Proceedings of the Twelfth Annual Conference on
  Computational Learning Theory}, pages 164--170. ACM Press, 1999.

\bibitem{6517175}
N.~Mohammed, D.~Alhadidi, B.~Fung, and M.~Debbabi.
\newblock Secure two-party differentially private data release for vertically
  partitioned data.
\newblock {\em Dependable and Secure Computing, IEEE Transactions on},
  11(1):59--71, Jan 2014.

\bibitem{Mohammed:2011:DPD:2020408.2020487}
N.~Mohammed, R.~Chen, B.~C. Fung, and P.~S. Yu.
\newblock Differentially private data release for data mining.
\newblock In {\em Proceedings of the 17th ACM SIGKDD International Conference
  on Knowledge Discovery and Data Mining}, KDD '11, pages 493--501, New York,
  NY, USA, 2011. ACM.

\bibitem{Narayan12djoin:differentially}
A.~Narayan and A.~Haeberlen.
\newblock Djoin: differentially private join queries over distributed
  databases.
\newblock In {\em In Proceedings of the 10th USENIX Symposium on Operating
  Systems Design and Implementation}, 2012.

\bibitem{Nissim:2007:SSS:1250790.1250803}
K.~Nissim, S.~Raskhodnikova, and A.~Smith.
\newblock Smooth sensitivity and sampling in private data analysis.
\newblock In {\em Proceedings of the Thirty-ninth Annual ACM Symposium on
  Theory of Computing}, STOC '07, pages 75--84, New York, NY, USA, 2007. ACM.

\bibitem{DBLP:conf/pods/PaghSW14}
R.~Pagh, M.~St{\"{o}}ckel, and D.~P. Woodruff.
\newblock Is min-wise hashing optimal for summarizing set intersection?
\newblock In {\em Proceedings of the 33rd {ACM} {SIGMOD-SIGACT-SIGART}
  Symposium on Principles of Database Systems, PODS'14, Snowbird, UT, USA, June
  22-27, 2014}, pages 109--120, 2014.

\bibitem{Paillier99public-keycryptosystems}
P.~Paillier.
\newblock Public-key cryptosystems based on composite degree residuosity
  classes.
\newblock In {\em Advances in Cryptology, EuroCrypt}, pages 223--238, 1999.

\bibitem{Poosala97selectivityestimation}
V.~Poosala.
\newblock Selectivity estimation without the attribute value independence
  assumption.
\newblock pages 486--495, 1997.

\bibitem{Sepehri12102014}
M.~Sepehri, S.~Cimato, and E.~Damiani.
\newblock Privacy-preserving query processing by multi-party computation.
\newblock {\em The Computer Journal}, 2014.

\bibitem{serfling1974}
R.~J. Serfling.
\newblock Probability inequalities for the sum in sampling without replacement.
\newblock {\em Ann. Statist.}, 2(1):39--48, 01 1974.

\bibitem{10.2307/2346966}
A.~B. Sunter.
\newblock List sequential sampling with equal or unequal probabilities without
  replacement.
\newblock {\em Journal of the Royal Statistical Society. Series C (Applied
  Statistics)}, 26(3):261--268, 1977.

\bibitem{Valiant84atheory}
L.~G. Valiant.
\newblock A theory of the learnable, 1984.

\bibitem{Vitter:1985:RSR:3147.3165}
J.~S. Vitter.
\newblock Random sampling with a reservoir.
\newblock {\em ACM Trans. Math. Softw.}, 11(1):37--57, Mar. 1985.

\bibitem{Woodruff_whendistributed}
D.~Woodruff and Q.~Zhang.
\newblock When distributed computation is communication expensive.
\newblock In Y.~Afek, editor, {\em Distributed Computing}, volume 8205 of {\em
  Lecture Notes in Computer Science}, pages 16--30. Springer Berlin Heidelberg,
  2013.

\bibitem{4568207}
A.~C.-C. Yao.
\newblock How to generate and exchange secrets.
\newblock In {\em Foundations of Computer Science, 1986., 27th Annual Symposium
  on}, pages 162--167, Oct 1986.

\end{thebibliography}

\end{document}